\documentclass[aps,preprint,floats,nofootinbib]{revtex4}
\usepackage{amsmath}
\usepackage{amssymb}
\usepackage{latexsym}
\usepackage{epsf}
\usepackage{graphicx}

\newlength{\defbaselineskip}
\newcommand{\setlinespacing}[1]%
           {\setlength{\baselineskip}{#1 \defbaselineskip}}

\def\lsim{\mathrel{\raise.3ex\hbox{$<$\kern-.75em\lower1ex\hbox{$\sim$}}}} 
\def\gsim{\mathrel{\raise.3ex\hbox{$>$\kern-.75em\lower1ex\hbox{$\sim$}}}} 
\begin{document}

\preprint{
\hfill
\begin{minipage}[t]{3in}
\begin{flushright}
\vspace{0.0in}
MADPH--05--1446\\
FERMILAB--PUB--05--434--A 
\end{flushright}
\end{minipage}
}

\hfill$\vcenter{\hbox{}}$

\vskip 0.5cm

\title{Prospects For Detecting Dark Matter With Neutrino Telescopes In Light Of Recent Results From Direct Detection Experiments}
\author{Francis Halzen$^1$ and Dan Hooper$^2$}
\address{$^1$Department of Physics, University of Wisconsin, Madison; \\$^2$Particle Astrophysics Center, Fermi National Accelerator Laboratory, Batavia, IL  60510-0500}

\date{\today}

\bigskip

\begin{abstract}

Direct detection dark matter experiments, lead by the CDMS collaboration, have placed increasingly stronger constraints on the cross sections for elastic scattering of WIMPs on nucleons. These results impact the prospects for the indirect detection of dark matter using neutrino telescopes. With this in mind, we revisit the prospects for detecting neutrinos produced by the annihilation of WIMPs in the Sun. We find that the latest bounds do not seriously limit the models most accessible to next generation kilometer-scale neutrino telescopes such as IceCube. This is largely due to the fact that models with significant spin-dependent couplings to protons are the least constrained and, at the same time, the most promising because of the efficient capture of WIMPs in the Sun. We identify models where dark matter particles are beyond the reach of any planned direct detection experiments while within reach of neutrino telescopes. In summary, we find that, even when contemplating recent direct detection results, neutrino telescopes still have the opportunity to play an important as well as complementary role in the search for particle dark matter.

\end{abstract}

\maketitle

\section{Introduction}

Many approaches have been developed to attempt to detect particles of dark matter. Such endeavors include direct detection experiments which hope to observe the scattering of dark matter particles with the target material of the detector and indirect detection experiments which are designed to search for the products of the annihilation of dark matter particles into gamma-rays, anti-matter and neutrinos \cite{neutrino}.

The sensitivity of direct detection experiments has been improving at a steady rate. The Cold Dark Matter Search (CDMS) experiment, operating in the Soudan mine in northern Minnesota, currently has produced the strongest limits on spin-independent scattering cross sections of WIMPs with nucleons \cite{cdmssi}, as well as on spin-dependent scattering cross sections of WIMPs with neutrons \cite{cdmssd}. The NAIAD experiment \cite{naiad} has placed the strongest constraints on spin-dependent WIMP-proton scattering. Several other experiments have placed limits that are only marginally weaker.

In addition to determining the rate in direct detection experiments, the elastic scattering cross sections of a WIMP also affects the sensitivity of high energy neutrino telescopes for indirect detection. Neutrino telescopes indirectly search for the presence of dark matter by taking advantage of the Sun's ability to capture large numbers of WIMPs over time. Over billions of years, a sufficiently large number of WIMPs can accumulate in the Sun's core to allow for their efficient annihilation. Such annihilations produce a wide range of particles, most of which are quickly absorbed into the solar medium. Neutrinos, on the other hand, may escape the Sun and be detected in experiments on the Earth. The prospects for such experiments detecting dark matter critically depend on the capture rate of WIMPs in the Sun, which in turn depends on the elastic scattering cross section of these particles. In this way, the sensitivity of indirect detection using neutrinos is coupled to the results of direct detection experiments \cite{earlyjungman}.\footnote{Neutrinos are also expected to be generated through dark matter annihilations in  the center of the Earth, although the prospects for detecting such neutrinos are very poor~\cite{edsjoearth}.}

The rapid progress of direct dark matter searches has been paralleled by the operation and development of large neutrino detectors. Currently, the Super-Kamiokande experiment has placed the strongest bounds on high-energy neutrinos from the direction of the Sun \cite{superk}. Super-K has two primary advantages over other experiments. Firstly, they have analyzed data over a longer period than most of their competitors, a total of nearly 1700 live days. Secondly, Super-K was designed to be sensitive to low energy ($\sim$GeV) neutrinos, which gives them an advantage in searching for lighter WIMPs. Super-K's limit on neutrino-induced muons above 1 GeV from WIMP annihilations in the Sun is approximately 1000 to 2000 per square kilometer per year for WIMPs heavier than 100 GeV, and approximately 2000 to 5000 per square kilometer per year for WIMPs in the 20 to 100 GeV range. The precise value of these limits depends on the WIMP annihilation modes considered.

The Amanda-II \cite{amanda}, Baksan \cite{baksan} and Macro \cite{macro} experiments have each placed limits on the flux of neutrino-induced muons from the Sun that are only slightly weaker than Super-Kamiokande's. The limit placed by the Amanda experiment resulted from only 144 live days of data. Having operated the detector for five years, Amanda is expected to produce significantly improved bounds in the future.

In addition to these experiments, the next generation neutrino telescopes IceCube and Antares are currently under construction at the South Pole and in the Mediterranean, respectively. IceCube, with a full cubic kilometer of instrumented volume, will be considerably more sensitive to WIMP annihilations in the Sun than other planned or existing experiments \cite{icecube}. Antares, with less than one tenth of the effective area of IceCube, will have the advantage of a lower energy threshold, and may thus be more sensitive to low mass WIMPs \cite{antares}.

In this article, we will attempt to assess the prospects for next generation neutrino telescopes to detect neutrinos generated in dark matter annihilations in the Sun, in light of the recent and projected constraints placed by direct detection experiments such as CDMS and NAIAD. We will begin by conducting a model-independent analysis, and will subsequently address specific dark matter models including the lightest neutralino in supersymmetric models and Kaluza-Klein states in models with Universal Extra Dimensions (UED). We find that, even when recent direct detection results are considered, there is an important and complementary role for neutrino telescopes to play in the search for particle dark matter.

\section{A Model Independent Assessment}

We will initially assume as little as possible about the particle nature of dark matter, for instance that it is a Weakly Interacting Massive Particle (WIMP) with a mass very roughly near the electroweak scale (GeV to several TeV). A generic species of WIMPs present in the solar system will scatter elastically with and become captured in the Sun at a rate given by~\cite{capture}
\begin{equation}
C^{\odot} \simeq 3.35 \times 10^{20} \, \mathrm{s}^{-1} 
\left( \frac{\rho_{\mathrm{local}}}{0.3\, \mathrm{GeV}/\mathrm{cm}^3} \right) 
\left( \frac{270\, \mathrm{km/s}}{\bar{v}_{\mathrm{local}}} \right)^3  
\left( \frac{\sigma_{\mathrm{H, SD}} +\, \sigma_{\mathrm{H, SI}}
+ 0.07 \, \sigma_{\mathrm{He, SI}}     } {10^{-6}\, \mathrm{pb}} \right)
\left( \frac{100 \, \mathrm{GeV}}{m_{\rm{WIMP}}} \right)^2 ,
\label{capture}
\end{equation}
where $\rho_{\mathrm{local}}$ is the local dark matter density, $\bar{v}_{\mathrm{local}}$ 
is the local rms velocity of halo dark matter particles and 
$m_{\rm{WIMP}}$ is the dark matter particle's mass. $\sigma_{\mathrm{H, SD}}$, $\sigma_{\mathrm{H, SI}}$ and $\sigma_{\mathrm{He, SI}}$ are the Spin Dependent (SD) and Spin Independent (SI) elastic scattering cross sections of the WIMP with hydrogen and helium nuclei, respectively. The factor of $0.07$ reflects the solar abundance of helium relative to hydrogen and well as dynamical factors and form factor suppression.

Notice that the capture rate is suppressed by two factors of the WIMP mass. One of these is simply the result of the depleted number density of WIMPs ($n \propto 1/m$) while the second factor is the result of kinematic suppression for the capture of a WIMP much heavier than the target nuclei, in this case hydrogen or helium. If the WIMP's mass were comparable to the masses of hydrogen or helium nuclei, these expressions would no longer be valid. For WIMPs heavy enough to generate neutrinos detectable in the high-energy neutrino telescopes, the result of Eq.~\ref{capture} should be applicable.

If the capture rate and annihilation cross sections are sufficiently large, equilibrium will be reached between these processes.  
For $N$ WIMPs in the Sun, the rate of change of this
quantity is given by
\begin{equation}
\dot{N} = C^{\odot} - A^{\odot} N^2  ,
\end{equation}
where $C^{\odot}$ is the capture rate and $A^{\odot}$ is the 
annihilation cross section times the relative WIMP velocity per volume. $A^{\odot}$ can be approximated by
\begin{equation}
A^{\odot} = \frac{\langle \sigma v \rangle}{V_{\mathrm{eff}}}, 
\end{equation}
where $V_{\mathrm{eff}}$ is the effective volume of the core
of the Sun determined roughly by matching the core temperature with 
the gravitational potential energy of a single WIMP at the core
radius.  This was found in Refs.~\cite{equ1,equ2} to be
\begin{equation}
V_{\rm eff} = 5.7 \times 10^{27} \, \mathrm{cm}^3 
\left( \frac{100 \, \mathrm{GeV}}{m_{\rm{WIMP}}} \right)^{3/2} \;.
\end{equation}
The present WIMP annihilation rate is given by
\begin{equation} 
\Gamma = \frac{1}{2} A^{\odot} N^2 = \frac{1}{2} \, C^{\odot} \, 
\tanh^2 \left( \sqrt{C^{\odot} A^{\odot}} \, t_{\odot} \right) \;, 
\end{equation}
where $t_{\odot} \simeq 4.5$ billion years is the age of the solar system.
The annihilation rate is maximized when it reaches equilibrium with
the capture rate.  This occurs when 
\begin{equation}
\sqrt{C^{\odot} A^{\odot}} t_{\odot} \gg 1 \; .
\end{equation}
If this condition is met, the final annihilation rate (and corresponding neutrino flux and event rate) has no further dependence on the dark matter particle's annihilation cross section.

As they annihilate, WIMPs can generate neutrinos through a wide range of channels. Annihilations to heavy quarks, tau leptons, gauge bosons and higgs bosons can all generate neutrinos in the subsequent decay. In some models, WIMPs can also annihilate directly to neutrino pairs. 

Once produced, neutrinos can travel to the Earth where they can be detected.  The muon neutrino spectrum at the Earth from WIMP annihilations in the Sun is given by:
\begin{equation}
\frac{dN_{\nu_{\mu}}}{dE_{\nu_{\mu}}} = \frac{ C_{\odot} F_{\rm{Eq}}}{4 \pi D_{\rm{ES}}^2}   \bigg(\frac{dN_{\nu}}{dE_{\nu}}\bigg)^{\rm{Inj}},
\label{wimpflux}
\end{equation}
where $C_{\odot}$ is the WIMP capture rate in the Sun, $F_{\rm{Eq}}$ is the non-equilibrium suppression factor ($\approx 1$ for capture-annihilation equilibrium), $D_{\rm{ES}}$ is the Earth-Sun distance and $(\frac{dN_{\nu}}{dE_{\nu}})^{\rm{Inj}}$ is the neutrino spectrum from the Sun per WIMP annihilating. Due to $\nu_{\mu}-\nu_{\tau}$ vacuum oscillations, the muon neutrino flux from WIMP annihilations in the Sun observed at Earth is the average of the $\nu_{\mu}$ and $\nu_{\tau}$ components. 

Muon neutrinos produce muons in charged current interactions with ice or water nuclei inside or near the detector volume of a high energy neutrino telescope. The rate of neutrino-induced muons observed in a high-energy neutrino telescope is estimated by: 
\begin{equation}
N_{\rm{events}} \simeq \int \int \frac{dN_{\nu_{\mu}}}{dE_{\nu_{\mu}}}\, \frac{d\sigma_{\nu}}{dy}(E_{\nu_{\mu}},y) \,R_{\mu}((1-y)\,E_{\nu})\, A_{\rm{eff}} \, dE_{\nu_{\mu}} \, dy,
\end{equation}
where $\sigma_{\nu}(E_{\nu_{\mu}})$ is the neutrino-nucleon charged current interaction cross section, $(1-y)$ is the fraction of neutrino energy which goes into the muon, $A_{\rm{eff}}$ is the effective area of the detector, $R_{\mu}((1-y)\,E_{\nu})$ is the distance a muon of energy, $(1-y)\,E_{\nu}$, travels before falling below the muon energy threshold of the experiment (ranging from $\sim$1 to 100 GeV), called the muon range. 

\begin{figure}[t]
\includegraphics[width=2.4in,angle=90]{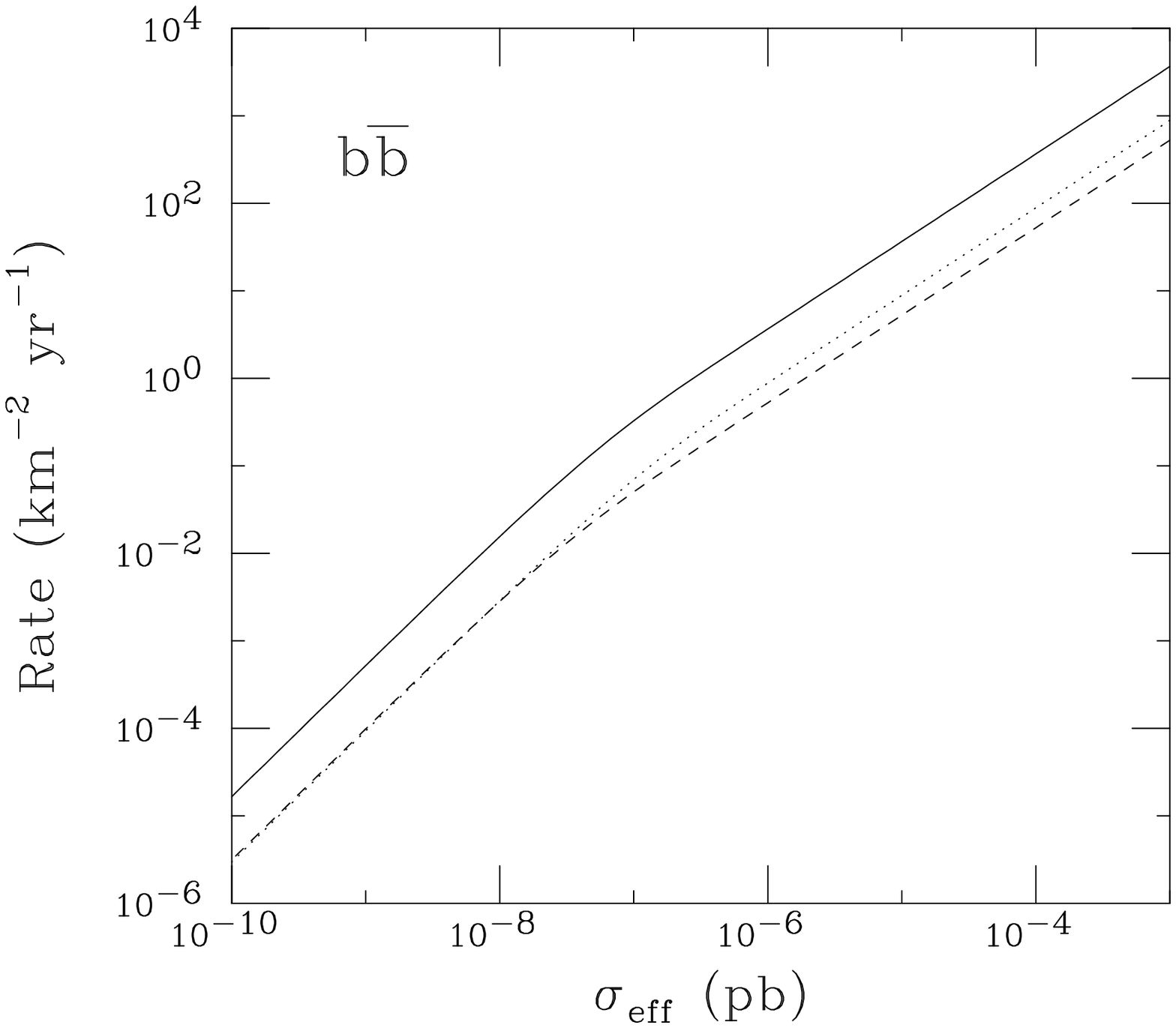}
\includegraphics[width=2.4in,angle=90]{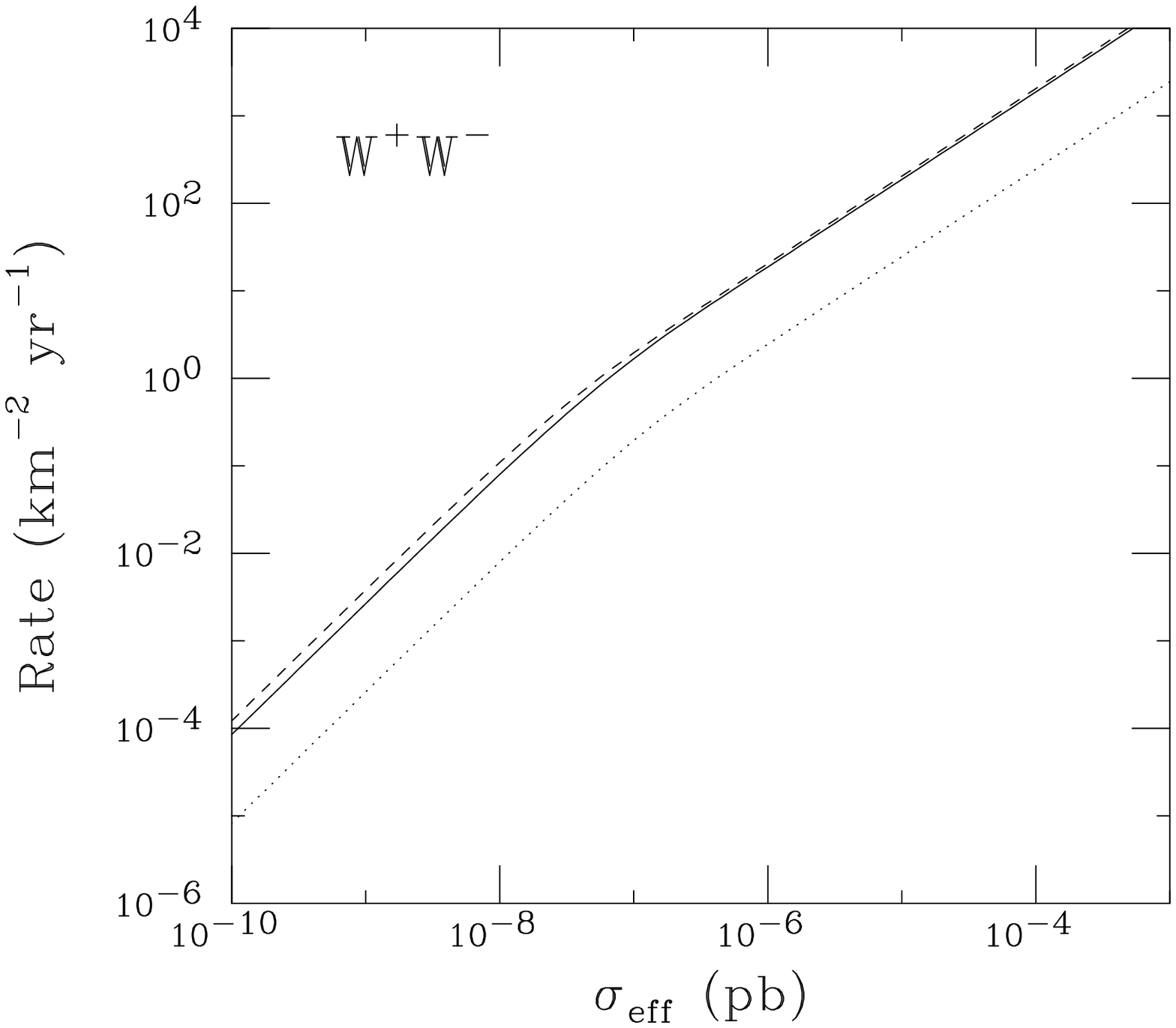}
 \\
\includegraphics[width=2.4in,angle=90]{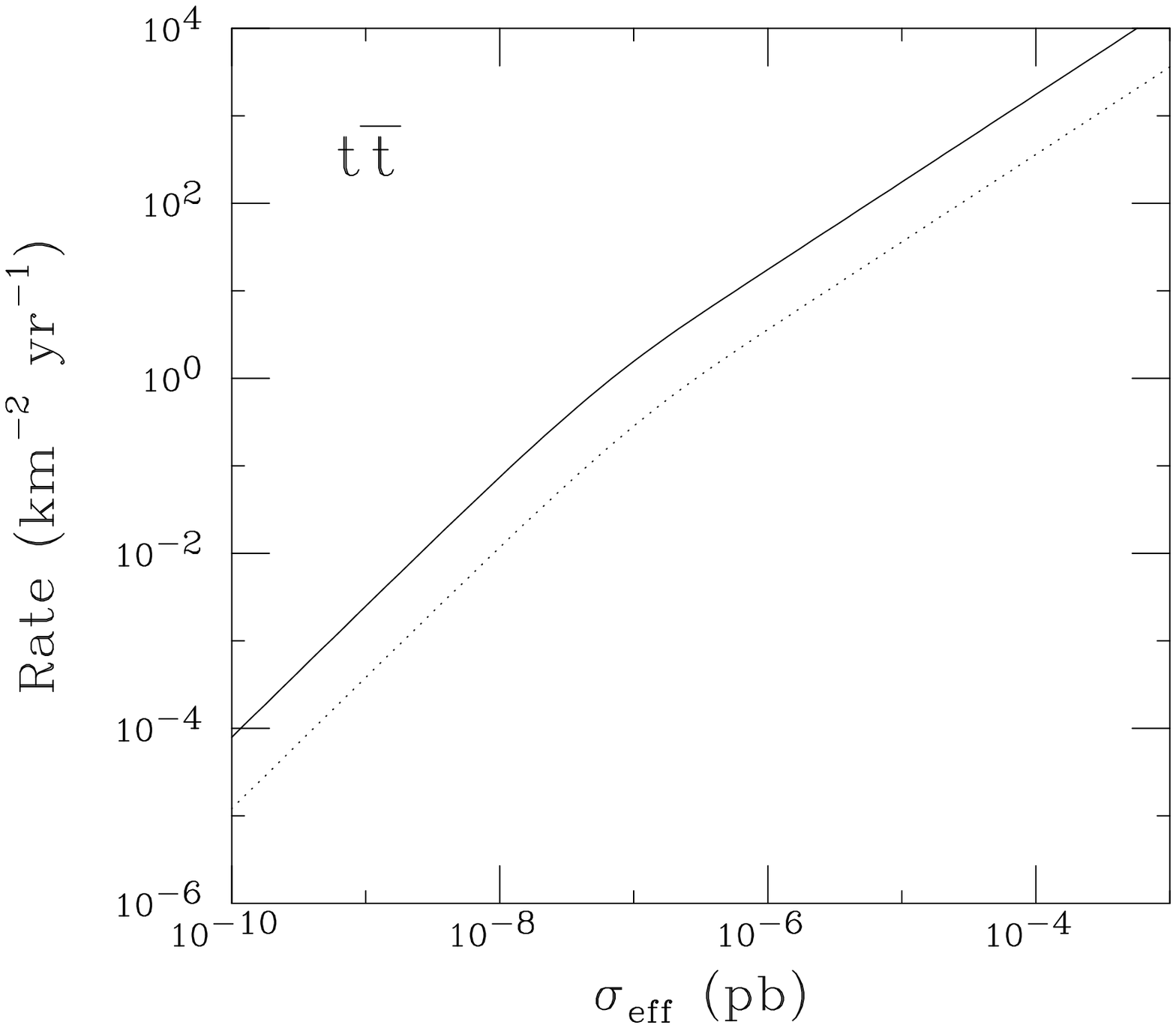}
\includegraphics[width=2.4in,angle=90]{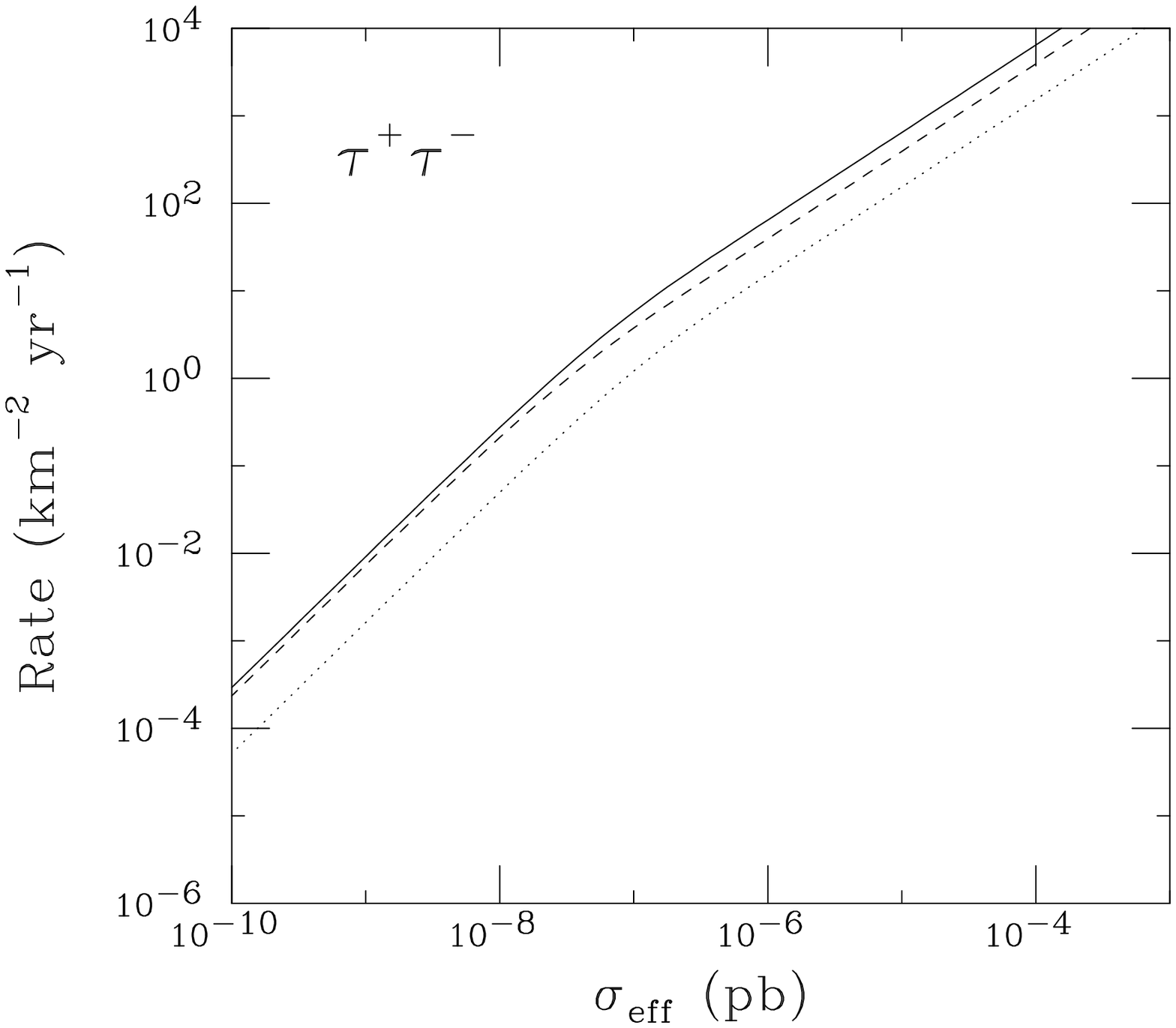}
 
\caption{The event rate in a kilometer-scale neutrino telescope as a function of the WIMP's effective elastic scattering cross section in the Sun for a variety of annihilation modes. The effective elastic scattering cross section is defined as $\sigma_{\rm{eff}} = \sigma_{\mathrm{H, SD}} +\, \sigma_{\mathrm{H, SI}}
+ 0.07 \, \sigma_{\mathrm{He, SI}}$, following Eq.~\ref{capture}. The dashes, solid and dotted lines correspond to WIMPs of mass 100, 300 and 1000 GeV, respectively. A 50 GeV muon energy threshold has been used. An annihilation cross section of $3 \times 10^{-26}$ cm$^-3$ s$^{-1}$ has been assumed. If another annihilation cross section were used, the change in the slope of these contours would occur at different location.}
\label{ratecompare}
\end{figure}

The spectrum and flux of neutrinos generated in WIMP annihilations depends on the annihilation modes which dominate, and thus is model dependent. As long as the majority of annihilations are to modes such as $b\bar{b}$, $t\bar{t}$, $\tau^+ \tau^-$, $W^+W^-$, $ZZ$, or some combination of higgs and gauge bosons, the variation from model to model is not dramaticl. In figure~\ref{ratecompare}, we plot the event rate in a kilometer-scale neutrino telescope as a function of the WIMP's effective elastic scattering cross section for a variety of annihilation modes. The effective elastic scattering cross section is defined as $\sigma_{\rm{eff}} = \sigma_{\mathrm{H, SD}} +\, \sigma_{\mathrm{H, SI}}
+ 0.07 \, \sigma_{\mathrm{He, SI}}$, following Eq.~\ref{capture}. 
The neutrino spectra were calculated following Ref.~\cite{Jungman:1994jr}, but including the effects of solar absorption according to the results of Ref.~\cite{crotty} an annihilation cross section of $3 \times 10^{-26}$ cm$^-3$ s$^{-1}$ has been assumed in the evaluation of capture-annihilation equilibrium.

For this calculation, and throughout this article, we will consider a kilometer-scale detector with a 50 GeV muon energy threshold. Such characteristics are meant to be represenative for either the IceCube experiment, or for a kilometer-scale neutrino telescope built in the Mediterranean sea. A 50 GeV muon energy threshold is perhaps somewhat optimistic for the design of IceCube, and somewhat conservative for a Mediterranean experiment with similar design to Antares. Detectors like IceCube can also implement, if motivated, strategies to lower the threshold, either in hardware or in trigger software. These are under study.

Notice that in figure~\ref{ratecompare}, the variation in the event rate in a neutrino telescope only modestly varies with the choice of dominant annihilation mode; within approximately one order of magnitude. In effect, the rate observed in a neutrino telescope acts as a determination of the effective elastic scattering cross section of WIMPs with protons and helium nuclei in the Sun. 

The elastic scattering cross section of a WIMP is constrained by the absence of a positive signal in direct detection experiments. Currently, the strongest limits on the WIMP-nucleon spin independent elastic scattering cross section have been made by the Cold Dark Matter Search (CDMS) experiment \cite{cdmssi}. This result excludes spin-independent cross sections larger than approximately $2 \times 10^{-7}$ pb for a 50-100 GeV WIMP or $7 \times 10^{-7}$ pb ($m_{\rm{WIMP}}$/500 GeV) for a heavier WIMP. The Zeplin-I \cite{zeplin} and Edelweiss \cite{edelweiss} experiments currently have spin-independent bounds that are roughly a factor of 5 weaker over this mass range.

With these results in mind, consider a 300 GeV WIMP with an elastic scattering cross section with nucleons which is mostly spin-independent. With a cross section near the CDMS bound, say $3 \times 10^{-7}$ pb, we can determine from figure~\ref{ratecompare} the corresponding rates in a kilometer-scale neutrino telescope, such as IceCube. Sadly, we find that this cross section yields only about 1 event per year for annihilations to $b\bar{b}$, 6 or 7 per year for annihilations to $W^+W^-$ or $t\bar{t}$ and about 20 per year for annihilations to $\tau^+ \tau^-$. Although 20 events per year from the Sun might be possible to distinguish from the atmospheric neutrino background, it would be a challenge. It is clear that WIMPs which scatter with nucleons mostly spin-independently are not likely to be detected with IceCube or other planned neutrino telescopes.

The same conclusion is not reached for the case of spin-dependent scattering, however. The strongest bounds on the WIMP-proton spin-dependent cross section have been made by the NAIAD experiment \cite{naiad}. This result limits the spin-dependent cross section with protons to be less than approximately 0.3 pb for a WIMP in the mass range of 50-100 GeV and less than 0.8 pb ($m_{\rm{WIMP}}$/500 GeV) for a heavier WIMP. The PICASSO \cite{picasso} and CDMS \cite{cdmssd} experiments have placed limits on the spin-dependent WIMP-proton cross section roughly one order of magnitude weaker than the NAIAD result. 

A WIMP with a largely spin-dependent scattering cross section with protons may thus be capable of generating large event rates in high energy neutrino telescopes. Again considering a 300 GeV WIMP with a cross section near the experimental limit, figure~\ref{ratecompare} suggests that rates as high as $\sim 10^6$ per year could be generated if purely spin-dependent scattering contributes to the capture rate of WIMPs in the Sun.

\section{The Case of Neutralino Dark Matter}

The lightest neutralino in R-parity conserving models of supersymmetry is, by far, the best studied candidate for dark matter. In this section, we consider the lightest neutralino of the Minimal Supersymmetric Standard Model (MSSM).

The elastic scattering and annihilation cross sections of a neutralino depend on its various couplings and on the mass spectrum of the higgs bosons and superpartners. The neutralino couplings depend on its composition. Generally, the lightest neutralino can be any mixture of bino, wino and the two CP-even higgsinos, although in most models a largely bino-like neutralino is lightest.

Spin dependent, axial-vector, scattering of neutralinos with quarks within a nucleon is made possible through the t-channel exchange of a $Z$, or the s-channel exchange of a squark. Spin independent scattering occurs at the tree level through s-channel squark exchange and t-channel higgs exchange, and at the one-loop level through diagrams involving a loop of quarks and/or squarks.

\begin{figure}[t]
\includegraphics[width=2.4in,angle=90]{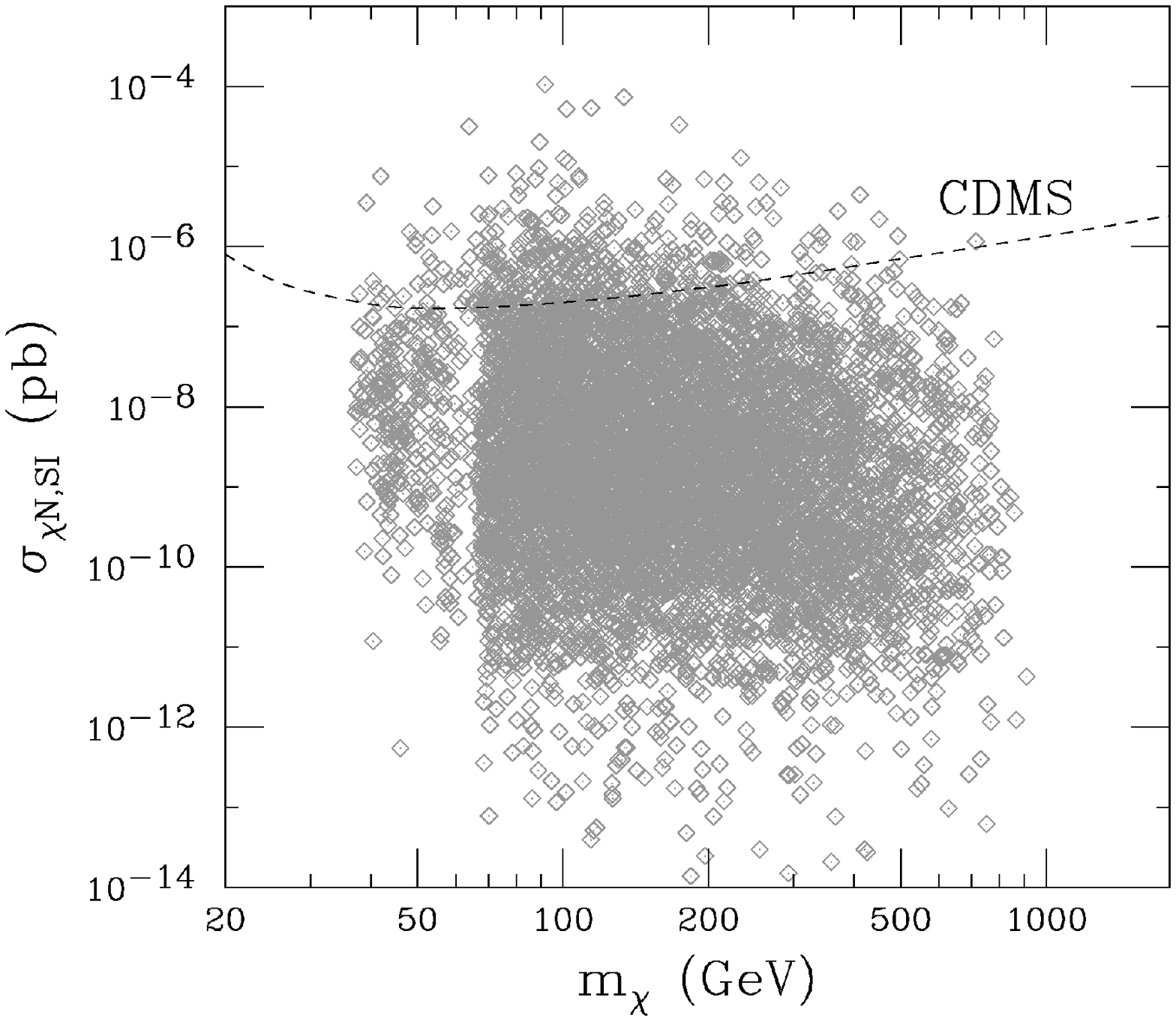}
\includegraphics[width=2.4in,angle=90]{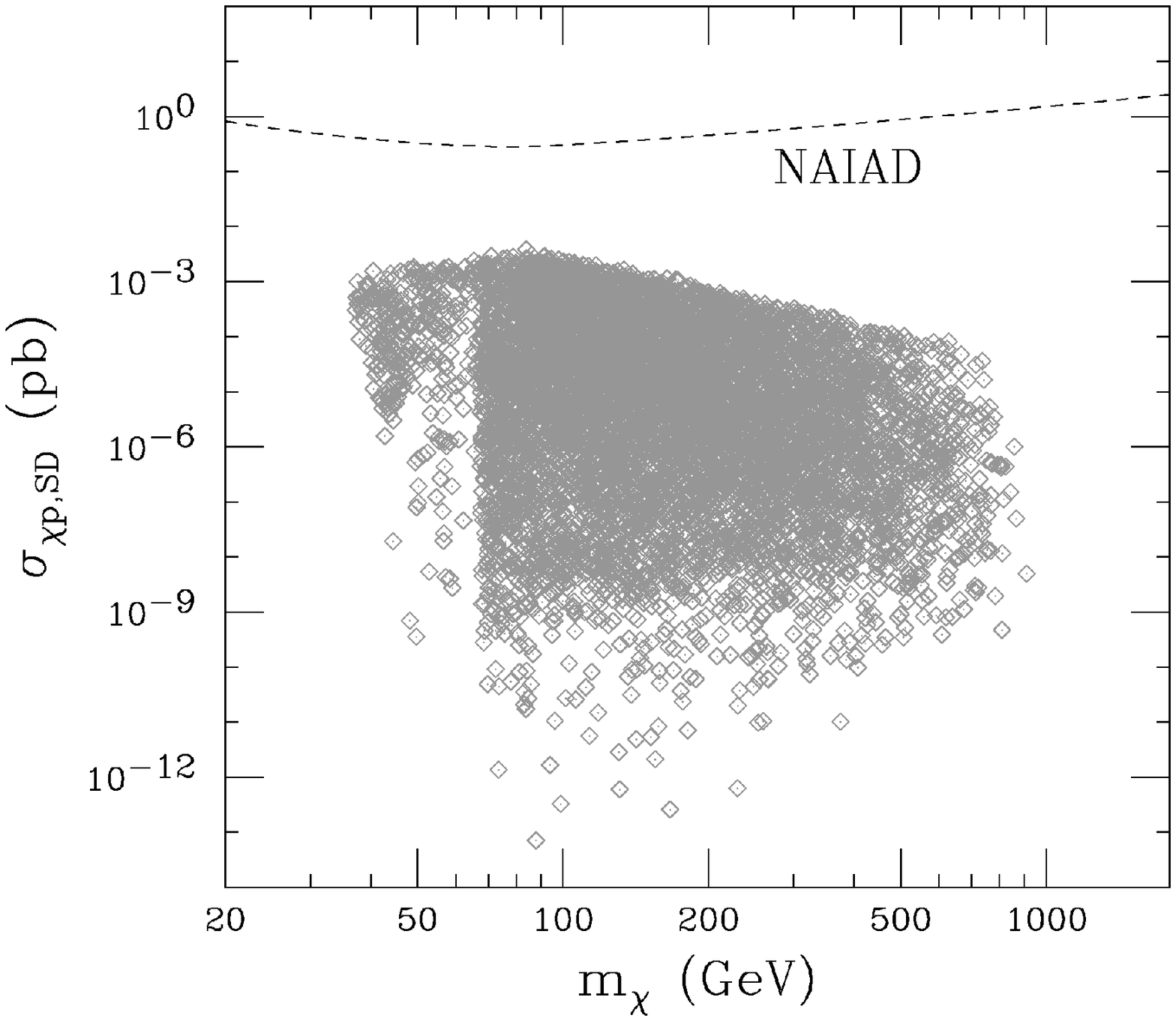}
 
\caption{The lightest neutralino's spin-independent (left) and spin-dependent (right) scattering cross sections for a range of MSSM parameters. See text for more details.}
\label{sigma}
\end{figure}

The cross sections for these processes can vary dramatically depending on the neutralino composition and the higgs and sparticle spectrum. In figure~\ref{sigma} we show the spin-dependent and independent scattering cross sections for a neutralino for a range of MSSM parameters. Our scan varied $M_1$, $M_2$, $M_3$, $\mu$ and all sfermion masses up to 10 TeV, $m_A$ up to 1 TeV and $\tan\beta$ between 1 and 60. For generality, we did not assume any particular SUSY breaking scenario or unification scheme. Each point shown is consistent with all collider constraints and does not produce a thermal relic density in excess of the value determined by WMAP, $\Omega_{\chi} h^2 \le 0.129$ \cite{wmap}. We do not impose a lower limit on this quantity, keeping in mind the possibility of non-thermal processes which may contribute to generating the density of neutralino dark matter. We have performed this scan using the DarkSUSY program \cite{darksusy}.

\begin{figure}[t]
\includegraphics[width=2.4in,angle=90]{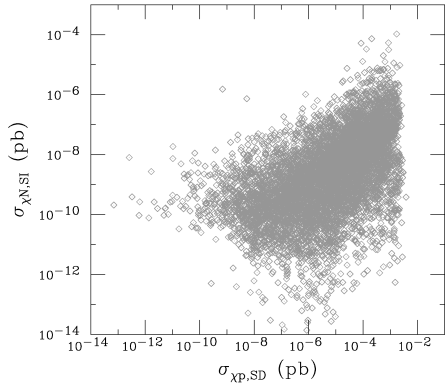}
\caption{A comparison of the spin-dependent and spin-independent scattering cross sections for neutralinos in the MSSM. See text for more details.}
\label{compare}
\end{figure}

\begin{figure}[t]
\includegraphics[width=2.4in,angle=90]{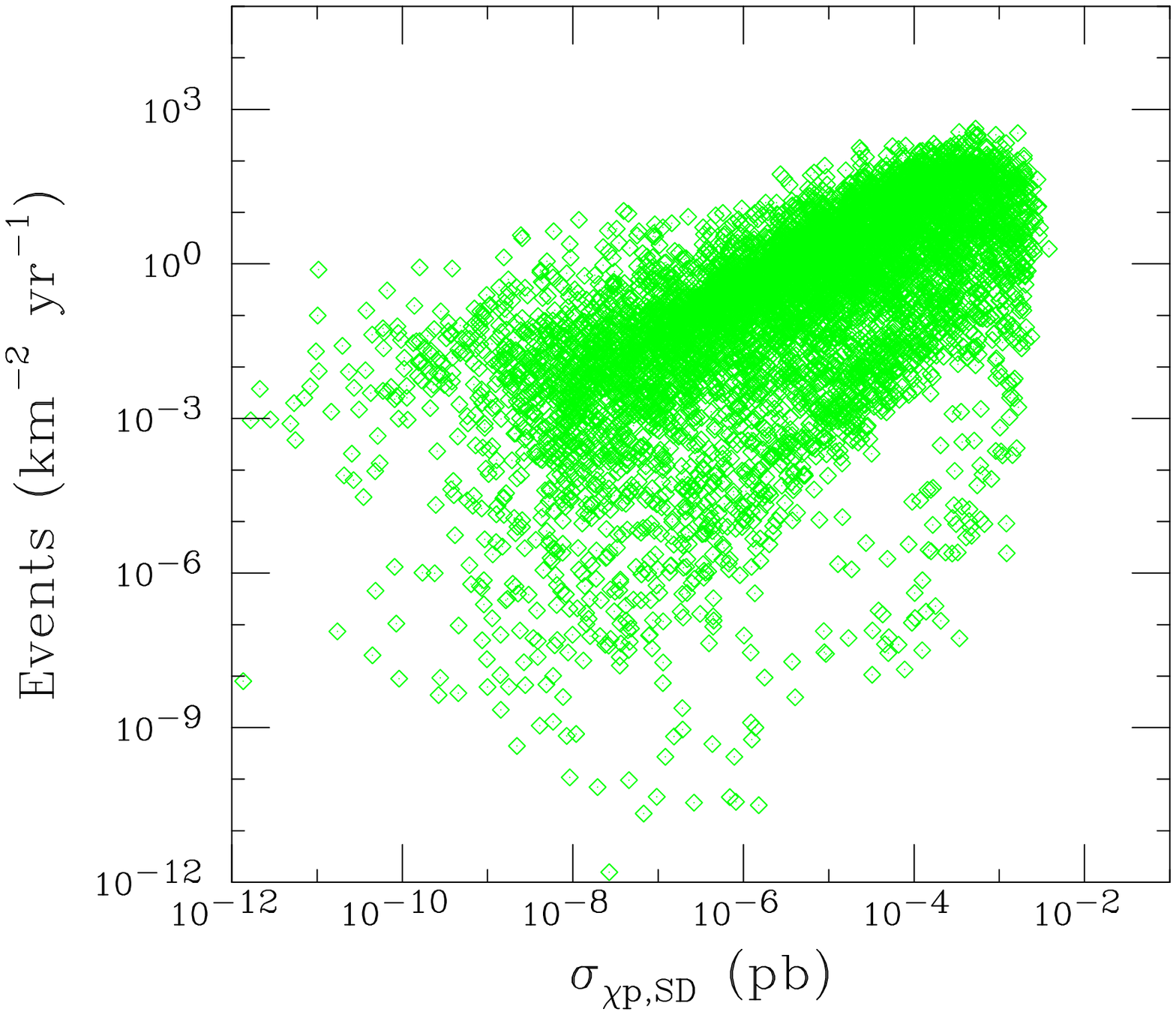}
\includegraphics[width=2.4in,angle=90]{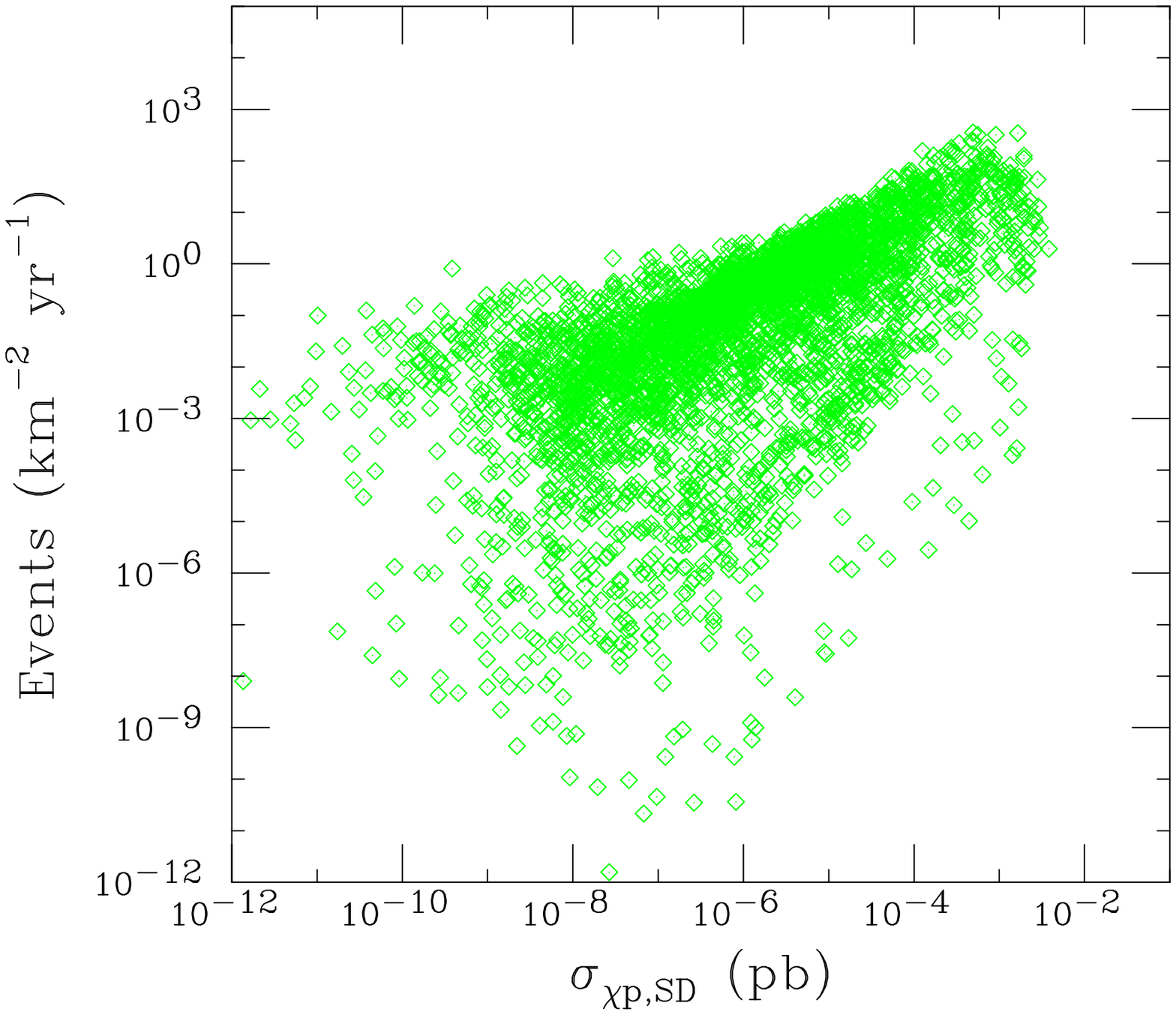}
\caption{The rate of events at a kilometer-scale neutrino telescope from dark matter annihilations in the Sun, as a function of the WIMP's spin-dependent elastic scattering cross section. In the left frame, no points shown violate the current spin-independent scattering constraints of CDMS. In the right frame, no points would violate the a spin-independent bound 100 times stronger. See text for more details.}
\label{ratesd}
\end{figure}

It is clear from figure~\ref{sigma} that for neutralinos, the spin dependent cross section can be somewhat larger than the spin independent, which is potentially well suited for the prospects for indirect detection. To further pursue this comparison, we plot in figure~\ref{compare} a direct comparison of these cross sections. We find that very large spin-dependent cross sections ($\sigma_{\rm{SD}} \gsim 10^{-3}$pb) are possible even in models with very small spin-independent scattering rates. Such a model would go easily undetected in all planned direct detection experiments, while still generating on the order of $\sim 1000$ events per year at IceCube. In figure~\ref{ratesd}, we demonstrate this by plotting the rate in a kilometer-scale neutrino telescope from WIMP annihilation in the Sun verses the WIMP's spin-dependent cross section with protons. In the left frame, all points evade the current constraints of CDMS. In the right frame, we plot the same result, but only showing those points which would evade a constraint {\it 100 times stronger} than the current CDMS bound. We thus conclude that next generation direct detection experiments will not be able to test all neutralino models accessible to an experiment such as IceCube.

\begin{figure}[t]
\includegraphics[width=2.4in,angle=90]{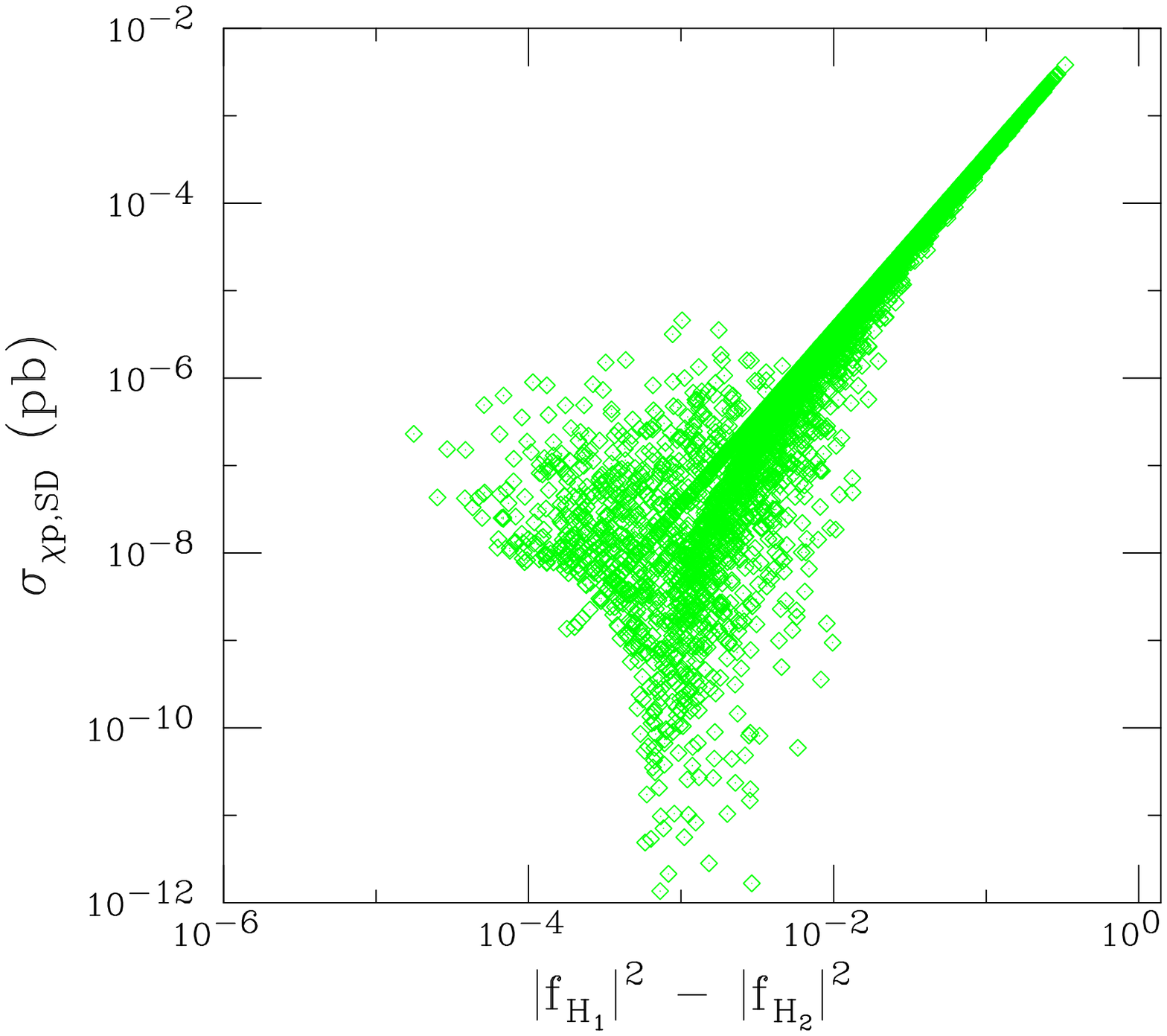}
\includegraphics[width=2.4in,angle=90]{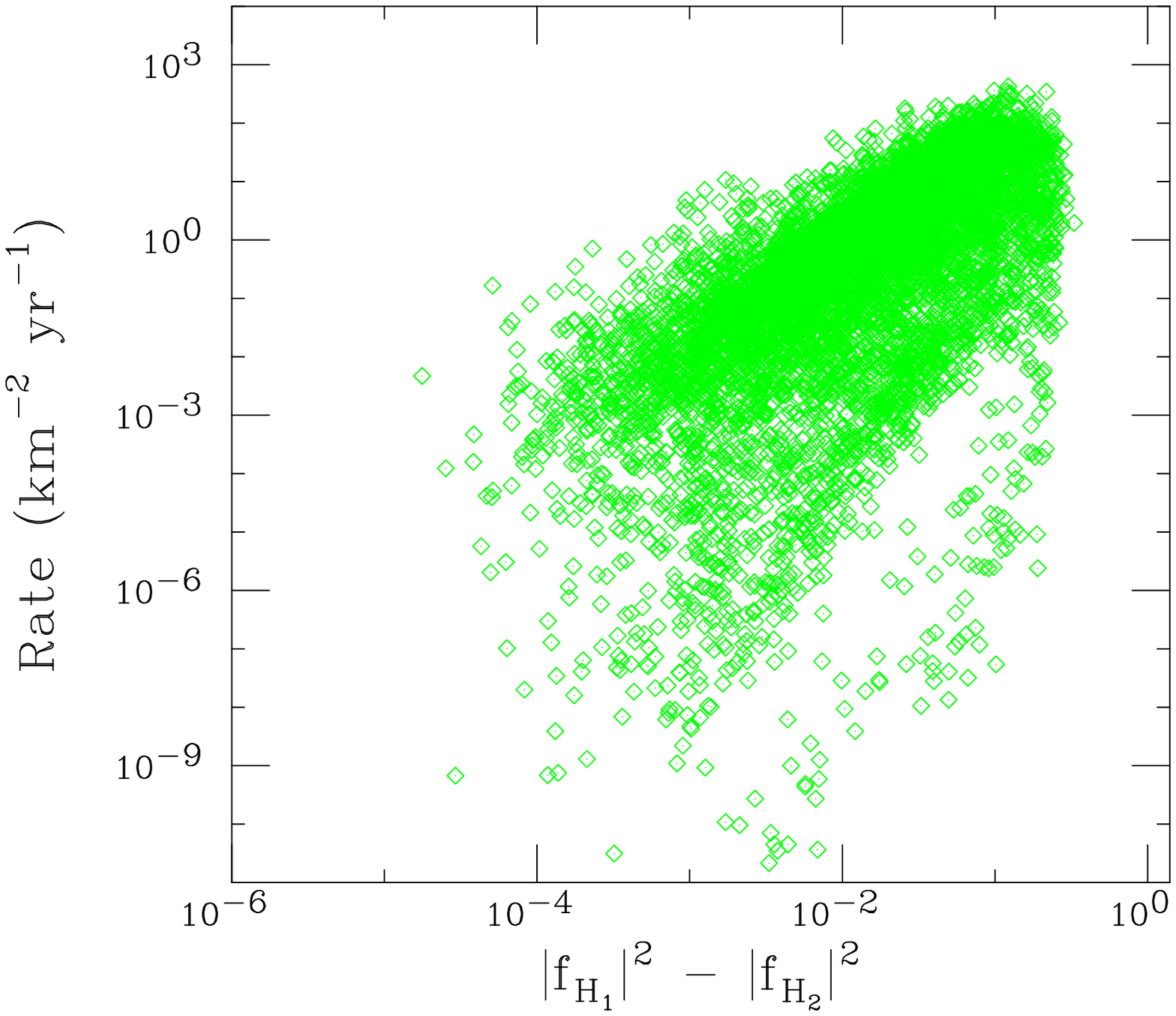}
\caption{In the left frame, the spin-dependent elastic scattering cross section of the lightest neutralino with protons is shown as a function of the quantity $|f_{H_1}|^2 - |f_{H_2}|^2$. In the right frame, the rate in a kilometer-scale neutrino telescope is shown, using a muon energy threshold of 50 GeV. Each point shown evades the current constraint of CDMS. See text for more details.}
\label{higgsino}
\end{figure}

A neutralino which has a large spin-dependent cross section generally has a sizable coupling to the $Z$, and thus has a large higgsino component. Writing the lightest neutralino as a superposition of the bino, wino and CP-even higgsino states, $\chi^0 = f_B \tilde{B} + f_W \tilde{W} + f_{H_1} \tilde{H_1} + f_{H_2} \tilde{H_2}$, the spin-dependent scattering cross section through the exchange of a $Z$ is proportional to the quantity $|f_{H_1}|^2 - |f_{H_2}|^2$. In figure~\ref{higgsino}, we plot this quantity verses the spin-dependent cross section, where this behavior becomes very evident. Neutralinos which are likely to be detectable in kilometer-scale neutrino telescopes are thus those with a higgsino component of approximately one percent or greater (or more precisely, $|f_{H_1}|^2 - |f_{H_2}|^2$ of approximately one percent or greater).

\section{The Case of Kaluza-Klein Dark Matter}

One alternative dark matter candidate which has received quite some attention recently arises in models with Universal Extra Dimensions (UED) \cite{kkdm}. In such models, all of the particles of the Standard Model can propagate through the bulk of the extra dimensional space, which is compactified on a scale around $R \sim \rm{TeV}^{-1}$. Each Standard Model particle is accompanied in this theory by a tower of Kaluza-Klein (KK) states. Such states appear as heavy versions of their Standard Model counterparts, with their extra mass being the result of momentum in the compactified dimensions of space. 

The Lightest KK Particle (LKP) in these models can be naturally stable, in a way analogous to how the lightest superpartner is stabilized in R-parity conserving models of supersymmetry. The most likely choice for the LKP is the first KK excitation of the hypercharge gauge boson, $B^{(1)}$. The thermal relic density of such a state can match the observed dark matter density over a range of masses, a few TeV $\gsim  m_{B^{(1)}} \gsim$ 500 GeV \cite{coann}. 

The elastic scattering cross sections for the LKP are quite challenging to reach with direct detection experiments, but are rather favorable for detection using neutrino telescopes. The spin-independent LKP-nucleon cross section, which is generated through the exchange of KK quarks and the higgs boson, is rather small and ranges between $10^{-9}$ and $10^{-12}$ pb \cite{kkelastic}, well beyond the range of current or upcoming direct detection experiments. The spin-dependent scattering cross section for the LKP with a proton, however, is considerably larger and is given by \cite{kkelastic}
\begin{equation}
\sigma_{H, SD} = \frac{g^{'4}\, m^2_p}{648 \pi m^4_{B^{(1)}} r^2_{q}} (4 \Delta_u^p + \Delta_d^p + \Delta_s^p)^2 \approx 4.4 \times 10^{-6} \,\rm{pb}\, \bigg(\frac{800 \,\rm{GeV}}{m_{B^{(1)}}}\bigg)^4 \, \bigg(\frac{0.1}{r_q}\bigg)^2,
\label{kkelastic}
\end{equation}
where $r_q \equiv (m_{q^{(1)}}-m_{B^{(1)}})/m_{B^{(1)}}$ is fractional shift of the KK quark masses over the LKP mass, which is expected to be roughly on the order of 10\%. The $\Delta$'s parameterize the fraction of spin carried by each variety of quark within the proton.

In addition to this somewhat large spin-dependent scattering cross section, the annihilation products of the LKP are very favorable for the purposes of generating observable neutrinos. Approximately 60\% of LKP annihilations generate a pair of charged leptons (20\% to each type). This is in contrast to neutralino annihilation in which the cross section to light fermion states is suppressed by a factor of $m^2_f/m^2_{\chi}$. Although most the remaining 40\% of LKP annihilations produce up-type quarks, about 4\% generate neutrino pairs. The neutrino and tau lepton final states each contribute substantially to the event rate in a neutrino telescope. 

\begin{figure}[t]
\includegraphics[width=2.4in,angle=90]{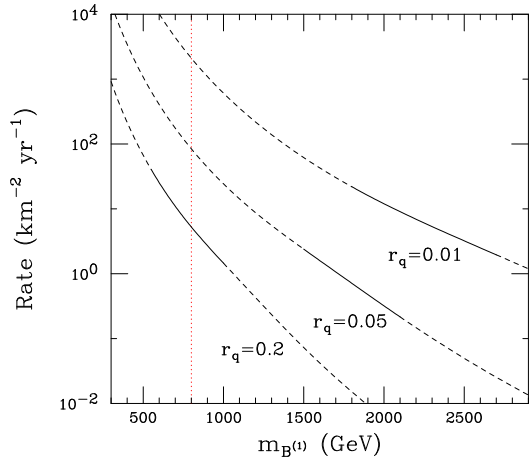}
\caption{The event rate in a kilometer-scale neutrino telescope as a function of the LKP mass. The three lines correspond to fractional mass splittings of the KK quarks of 20\%, 5\% and 1\%. The solid sections of these lines reflect the approximate range in which it is possible to generate the observed thermal relic abundance, following Ref.~\cite{coann}. Masses to the left of the vertical dotted line are excluded at the 95\% confidence level by electroweak precision observables measured at LEP 1 and LEP 2~\cite{ew}.  A 50 GeV muon energy threshold has been used.}
\label{kkrates}
\end{figure}

The event rates in a kilometer scale neutrino telescope from KK dark matter annihilating in the Sun are estimated in figure~\ref{kkrates}. There are competing effects which contribute to these results. In particular, a small mass splitting between the LKP and KK quarks yields a large spin-dependent elastic scattering cross section, as seen in Eq.~\ref{kkelastic}. On the other hand, KK quarks which are not much heavier than the LKP contribute to the freeze-out process and increase the range of LKP masses in which the thermal abundance matches the observed dark matter density. Over the range of masses which the observed dark matter density can be thermally generated (based on the calculations of Ref.~\cite{coann}), which are shown as the solid line segments in figure~\ref{kkrates}, between roughly 0.5 and 10 events per year are expected in a kilometer scale neutrino telescope such as IceCube. If non-thermal mechanisms, such as decays of KK gravitons, contribute to generating the LKP relic abundance, much larger rates may be possible.

\section{Discussion and Conclusions}
\label{conc}

In this article, we have revisted the prospects for the detection of dark matter annihilating in the core of the Sun, using neutrino telescopes. This is an issue of particular relevance given the recent advancements being made in the efforts of direct detection.

Although the prospects for the indirect detection of dark matter with high-energy neutrinos are not independent to the results of direct detection experiments, this relationship is not as straightforward as it is often assumed. We have shown, in particular, that the latest bounds placed on the scalar WIMP-nucleon cross section by the CDMS experiment do not seriously limit the models most accessible to next generation kilometer-scale neutrino telescopes such as IceCube. This is largely due to the fact that the spin-dependent scattering of WIMPs with protons is the most efficient process for capture in the Sun for many particle dark matter models. The spin-dependent scattering cross section of a WIMP is not nearly as strongly constrained as the spin-independent quantity.

We have explored two specific models for the particle nature of dark matter in this article: neutralinos in supersymmetric models, and Kaluza-Klein states in UED models.  We find that the best prospects for the detection of neutralino dark matter with neutrino telescopes are in those models in which the LSP has a considerable higgsino fraction ($\sim1$\% or more). For dark matter in the form of the first Kaluza-Klein excitation of the hypercharge gauge boson, $B^{(1)}$, we find that substantial rates are generically anticipated.

We would now like to take the time to comment on the prospects of detecting other possible particle candidates for WIMP dark matter. As we have emphasized, the key feature needed in such a candidate for it to be detected by a neutrino telscope is a large spin-dependent scattering cross section with protons. Other characteristics can also contribute, however. Annihilations to modes such as tau leptons, gauge bosons and directly to neutrinos have an over all positive effect on the anticipated event rates. WIMP masses a factor of roughly 4-6 above the muon energy threhsold of the experiment being considered is ideal. Below this range, substantial numbers of muons have too little energy to be detected, while at higher masses, fewer neutrinos are generated, and solar absorption is more efficient.

There are also astrophysical considerations which might effect our conclusions, although modestly. In particular, if the local dark matter distribution is not smooth and homogeneous, direct detection experiments may be currently sampling a density of dark matter which is above or below the average value. The capture of dark matter in the Sun, in contrast, has been averaged over several billions of years, and is far less strongly effected by such density fluctuations. As CDMS and other direct detection experiments actually constraint the product of the WIMP's scattering cross section and the local dark matter density, it could be imagined that these bounds might be altered by a factor of a few due to local dark matter inhomogeneities.

Finally, to make a point that exceptions may exist to the conclusions we draw here, we will briefly consider one other particle dark matter candidate. In particular, we will discuss a sneutrino LSP which is a mixture of a sneutrino and the superpartner of a sterile state. Such a model can be arranged such that the sneutrino LSP does not elastically scatter efficiently with nucleons, but can scatter inelastically into a state on the order of 100 keV more massive \cite{smithweiner}. This leads to rather suppressed rates in direct detection experiment, without dramatically reducing the capture rate in the Sun. In this way, neutrino fluxes as large or larger than the current bounds placed by Super-Kamionkande, Macro, Baksan and Amanda can be generated without exceeding direct detection limits. Such a scenario has been explored in an effort to reconcile the results of DAMA with CDMS and other direct detection experiments \cite{smithweiner}. This model is also capable of providing a natural explanation for the observed ratio of the dark matter and baryonic matter densities \cite{sneutrinomarch}.

\begin{acknowledgements}
DH is supported by the US Department of Energy and by NASA grant NAG5-10842.
\end{acknowledgements}

\end{document}